\documentclass[twocolumn,prl,aps]{revtex4}

\usepackage{graphicx}
\usepackage{dcolumn}
\usepackage{bm}
\usepackage{psfig}
\usepackage{epsfig}

\def\la{\langle}
\def\ra{\rangle}

\def\k2av{\la k_T^2\ra}

\def\htm{\hat{t}}

\newcommand{\beq}{\begin{equation}}
\newcommand{\beqar}{\begin{eqnarray}}
\newcommand{\eeq}[1]{\label{#1} \end{equation}}
\newcommand{\eeqar}[1]{\label{#1} \end{eqnarray}}
\begin{document}
\title{\bf Kaon Fragmentation Function and $K/\pi$ Ratios 
in Nuclear Collisions}
\author{Yi Zhang}
\address{Mathematics Department, 
         Kent State University, Kent, OH 44242}

\begin{abstract}
An improved leading order fragmentation functions set of kaon 
is proposed based on the experimental data. We compare it with 
currently available sets, and use it to calculate high-$p_T$ 
$K/\pi$ ratios in relativistic proton-proton collisions. A 
prediction of $K/\pi$ ratios in relativistic nucleus-nucleus
collisions ($d + Au$ and $Au + Au$) at RHIC $\sqrt{s} = 200$ GeV 
is given. 
\end{abstract}
\pacs {PACS number(s): 12.38.Bx, 13.85.Ni, 13.85.Qk, 24.85.+p, 25.75.-q}
\maketitle

An enhancement of the abundance of strange quarks ($s$ and 
$\bar{s}$) is expected if a quark-gluon plasma (QGP) is formed
in a relativistic nuclear collision, compared to a collision 
without QGP formation\cite{Rafel82}. This asymmetry in the flavor
composition of QGP may be reflected in the final particle
composition. This is therefore of interest to systematically study 
the production of strange mesons, in particular kaons, in relativistic 
proton-proton ($pp$) and proton-nucleus ($pA$) collisions. Furthermore, 
it has been suggested~\cite{Peter00} that the measurable $K/\pi$ ratios 
are sensitive to the initial density of the QGP in relativistic
nucleus-nucleus ($AB$) collisions. Henceforth, in order to provide  
a basis for the treatment of kaon production in nuclear collisions
at the energies of the Relativistic Heavy Ion Collider (RHIC),
it is important and necessary to first develop a good description of  
kaon production in $pp$ collisions and relying on that, we can then 
give a reliable studying of hard (high-$p_T$) $K/\pi$ probe for nuclear 
collisions.

\begin{figure}[t!]
\vspace*{0.5cm}
\epsfig{file=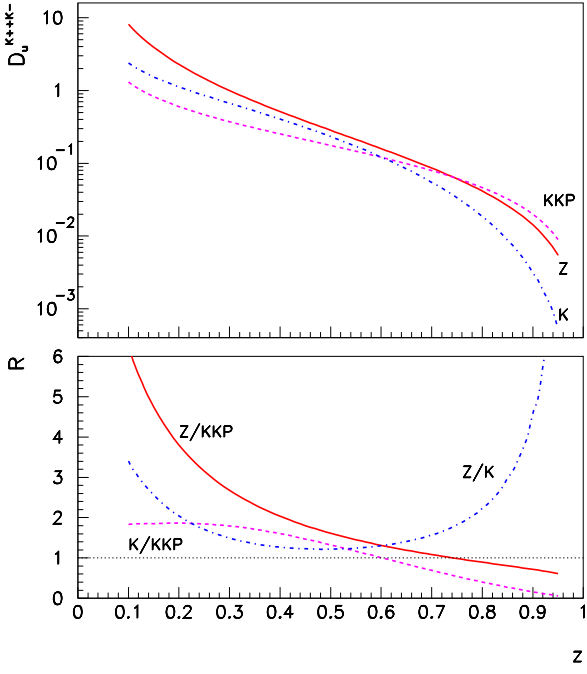,height=3.9in,
width=3.4in,clip=5,angle=0}
\vspace*{0.0cm}
\caption[]{
\label{fig:fig1}
The comparisons of improved FFs (full line) with KKP-FFs (dashed line) 
and K-FFs (dot-dashed line).
In top panel we show the fragmentation function of $u$ quark to
$K^+ + K^-$, $D_{u}^{K^+ + K^-}$. In the bottom panel, the ratios 
between different FFs sets, Z/KKP (full line), K/KKP (dashed
line), and Z/K (dot-dashed line), are shown, as a function of $z$ 
at fixed scale $Q^2 = 25$ $GeV^2$.}
\end{figure}

In the framework of parton model of perturbative QCD (pQCD), the 
invariant cross section for inclusive kaon production in a $pp$ 
collision gets involved with fragmentation functions (FFs), 
$D_{h/a} (z,Q^2)$, through the factorization theorem~\cite{Sterman96}.
The value of $D_{h/a} (z,Q^2)$ corresponds to the probability for
the parton $a$ produced at short distance $\sim 1/Q$ to form a jet
that includes the hadron $h$ carrying the fraction $z$ of the
longitudinal momentum of parton $a$. It is a nonperturbative 
object and hence can not be derived from QCD. During the last 
few years, the experimental progress at the CERN and the SLAC 
has motivated people to generate leading-order (LO) and 
next-to-leading-order (NLO) set of
$\pi^\pm$, $K^\pm$, and $p/\bar p$ FFs set~\cite{BKK,KKP,Kre},
which can be utilized in pQCD calculations. However, the 
available FFs sets for kaon are still not satisfied. For example, 
in Fig.\ref{fig:fig1}, we show the comparison between Kretzer's 
FFs (K) and that of Kniehl, Kramer, and P{\"o}tter (KKP),
for the fragmentation process of $u$ quark to $K^+ + K^-$.
We find that the differences between KKP-FFs and K-FFs are significant,
especially at large-$z$ region.

On the other hand, the experimental data of kaon production in $pp$ 
collisions in the energy range 20 GeV 
$\lesssim \sqrt{s} \lesssim$ 40~GeV\cite{antreasyan79,E605pi} shows
that $K^+/\pi^+ \rightarrow 1/2$ at high-$p_T$ region, which reminds 
us that $D_u^{K^+}/D_u^{\pi^+} \rightarrow 1/2$. While interesting
enough, it has been implied by Feynman {\it et al.} that, when 
$z \rightarrow 1$ ($z$ is the longitudinal momentum fraction of a
parton), $D_u^{K^+}/D_u^{\pi^+} \rightarrow (1-\beta)/\beta$, in   
which $\beta$ is a constant parameter in Feynman-Field 
parameterization\cite{Feynman,Field}. In current letter, based on the
KKP-FFs of pion and experimental data, we propose the following improved 
LO FFs (Z) $D_{Z}^{K^\pm}$ of kaon,
\begin{equation}
\label{kff}
D_{a,Z}^{K^\pm} = \frac{1}{2} D_{a,KKP}^{\pi^\pm} \ , 
\end{equation}
in which $D_{a,KKP}^{\pi^\pm}$ is the corresponding LO KKP-FFs of pion 
and parton $a$ can be the $u (\bar u)$, $d (\bar d)$, $s (\bar s)$ quark 
or gluon ($g$).

To deal with charged kaons, i.e., $K^+$  and $K^-$, we now consider
the scheme to separate the above FFs for kaon. A natural way to do
is as follows: we assume, say, for $u$ valence quark of $K^+ (u\bar s)$,
\begin{equation}
\label{sff1}
D_{u,Z}^{K^+} = \frac{1}{2} D_{u,KKP}^{\pi^+} \ ,
\end{equation}
so that we have 
\begin{equation}
\label{sff2}
D_{u,Z}^{K^-} = 0 \ , 
\end{equation}
which means the contribution of $u$ valence quark to $K^- (\bar u s)$ 
is vanished.

Starting with eqn-s.(\ref{sff1}) and (\ref{sff2}), we compare our 
improved Z-FFs with KKP-FFs and K-FFs in Fig.\ref{fig:fig1}. 
We can see that the Z-FFs agrees with K-FFs in the most of moderate 
$z$ region, but disagrees with it at large-$z$ region (with $z > 0.7$), 
where in stead tends to agree with KKP-FFs. 

Based on this Z-FFs set, the LO invariant cross section of kaon 
production in $pp$ collisions is described in the pQCD-improved parton
model on the basis of the factorization 
theorem as a convolution~\cite{Field}:
\begin{eqnarray}
\label{hadk}
 && E_K \frac{d \sigma^{LO}_{pp}}{d^3 p} =
 \sum_{abcd} \int dx_{a}\, dx_{b}\, d{\bf k}_{Ta}\, d{\bf k}_{Tb}\, dz_c
   \  \nonumber \\
 && \ \ \ \times 
 g_{a/p}({\bf k}_{Ta}) \, f_{a/p}(x_a, Q^2)\, g_{b/p}({\bf k}_{Tb})\, f_{b/p}(x_b, Q^2)
   \  \nonumber \\
 && \ \ \ \times
 \frac{d\sigma^{LO}}{d\hat{t}}(ab \to cd) \frac{D^{Z}_{K/c}(z_c, Q^{'2})}{\pi z_c^2} \hat{s} \delta(\hat{s}+\hat{t}+\hat{u}) \,,
\end{eqnarray}
where $f_{a/p}(x,Q^2)$ and $f_{b/p}(x,Q^2)$ are the parton distribution
functions (PDFs) for the
colliding partons $a$ and $b$ in the interacting protons
as functions of momentum fraction $x$, at scale $Q$,
$d \sigma/ d \htm$ is the hard scattering cross section of the
partonic subprocess $ab \to cd$, and the FFs, $D^{Z}_{K/c}(z_c,Q'^2)$
gives the probability for parton $c$ to fragment into kaon
with momentum fraction $z_c$ at scale $Q'$. The superscript ''Z" is used
here for our improved Z-FFs. We use the convention
that the parton-level Mandelstam variables are written with a `hat' (like
$\htm$ above). The scales are fixed in the present work as $Q = p_T/2$
and $Q' = p_T/2z_c$. The function $g_{a/p}({\bf k}_T)$ is the intrinsic 
transverse momentum distributions of parton $a$ inside nucleon.
We follow the phenomenological approach
in the present work, taking $g({\bf k}_T)$ to be a Gaussian:
\begin{equation}
\label{kTgauss}
g({\bf k}_T) \ = \frac{1}{\pi \la k^2_T \ra}
        e^{-{k^2_T}/{\la k^2_T \ra}}    \,\,\, .
\end{equation}
Here $\langle k_T^2 \rangle$ is the 2-dimensional width of the $k_T$
distribution and it is related to the magnitude of the
average transverse momentum of one parton
as $\langle k_T^2 \rangle = 4 \langle k_T \rangle^2 /\pi$.

In this investigation we use LO partonic cross sections, together
with LO PDFs (GRV)~\cite{GRV92}. The Monte-Carlo integrals are 
carried out by the VEGAS-routine~\cite{VEGAS}. 

We analysed experimental data of inclusive $K$ production in 
$pp$ collisions in the center-of-mass energy range
20 GeV $\lesssim \sqrt{s} \lesssim$ 40~GeV\cite{antreasyan79,E605pi}.
We find a good agreement between our model and experimental data on
both $K^+/\pi^+$ and $K^-/\pi^-$. In our analysis, the intrinsic
transverse momentum distribution of parton, $\langle k_T^2 \rangle =$
1.7, 2.0, 2.2, and 2.2 $GeV^2$, are used for both kaon and pion at 
$\sqrt{s} = $ 19.4, 23.7, 27.4, and 38.8 GeV, respectively~\cite{zfpgl02}.
   
\begin{figure}[t!]
\vspace*{0.5cm}
\epsfig{file=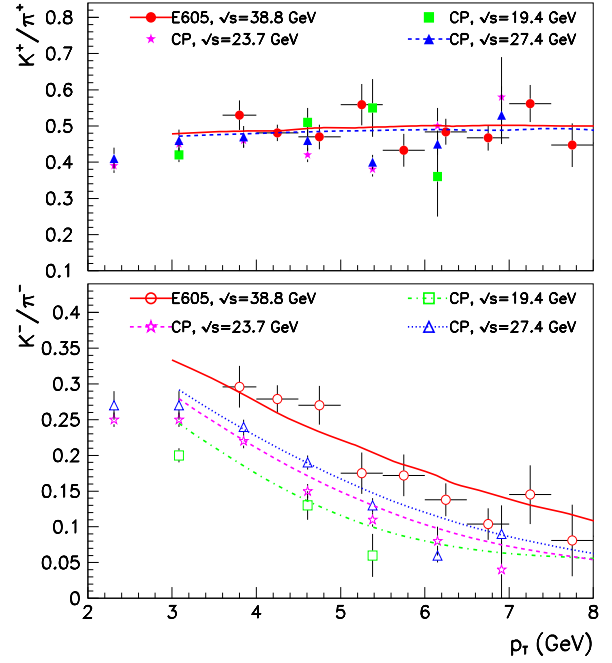,height=3.9in,
width=3.4in,clip=5,angle=0}
\vspace*{0.0cm}
\caption[]{
\label{fig:fig2}
The comparison of experimental (symbols) and pQCD-calculated (curves)
$K/\pi$ ratios: $K^+/\pi^+$ (top panel) and $K^-/\pi^-$ (bottom panel), 
as a function of $p_T$ at $\sqrt{s} = $ 19.4, 23.7, 27.4, and 38.8 GeV, 
respectively, in $pp$ collisions. In top panel, we only show $K^+/\pi^+$ 
ratios at $\sqrt{s} = $ 27.4 (full line) and 38.8 (dashed line) GeV. 
The kaon production is calculated based on improved Z-FFs. The experimental
data of $K/\pi$ ratios are from Refs.\cite{antreasyan79,E605pi}.} 
\end{figure}

High-$p_T$ kaon production data begun to emerge from RHIC, at
center-of-mass energies $\sqrt{s} = $ 130 and 200 GeV~\cite{QM01, QM02}. 
This provides us with the opportunity to predict kaon production
in $d+Au$ and $Au+Au$ collisions at higher energies. 
For nucleus-nucleus collisions, the Glauber model\cite{Glauber}
is the standard description. The LO inclusive kaon production in $AB$ 
collisions can be written in the following formlism:
\begin{eqnarray}
\label{AAK}
 && E_K \frac{d \sigma^{LO}_{AB}}{d^3 p} =
 \sum_{abcd} \int d^2 bd^2 r{\it T}_A(r){\it T}_B(|{\bf b}-{\bf r}|)\int dx_{a} \  \nonumber \\
 && \ \ \ \ \ \times
 \,dx_{b} d{\bf k}_{Ta} d{\bf k}_{Tb}\,dz_c\,\, g_{a/A}({\bf k}_{Ta})f_{a/A}(x_a, Q^2) 
   \  \nonumber \\
 && \ \ \ \ \ \times
 \, g_{b/B}({\bf k}_{Tb})\, f_{b/B}(x_b, Q^2)\, \frac{d\sigma^{LO}}{d\hat{t}}(ab \to cd)
   \  \nonumber \\
 && \ \ \ \ \ \times
 \,\, \frac{D^{Z}_{K/c}(z_c, Q^{'2})}{\pi z_c^2} \hat{s} \delta(\hat{s}+\hat{t}+\hat{u}).
\end{eqnarray}
 
\begin{figure}[t!]
\vspace*{0.5cm}
\epsfig{file=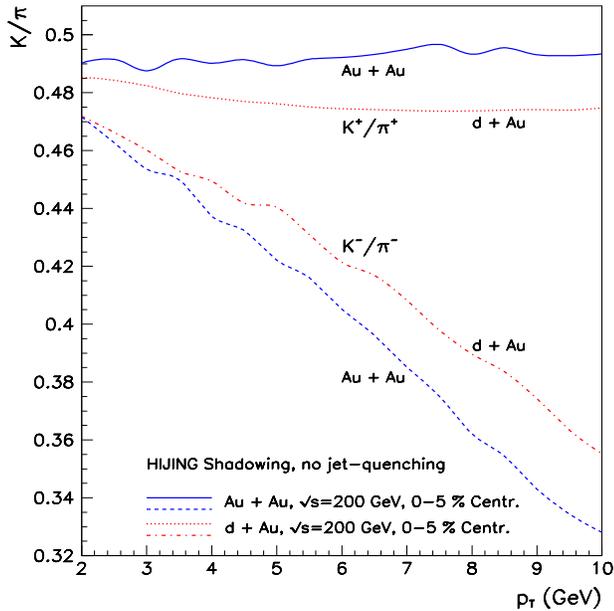,height=3.6in,
width=3.5in,clip=5,angle=0}
\vspace*{0.0cm}
\caption[]{
\label{fig:fig3}
The predicted $K/\pi$ ratios for $d + Au$ and $Au + Au$ collisions
at RHIC energy $\sqrt{s} = 200$ GeV (0 - 5\% centrality) with HIJING 
shadowing, but without 
jet quenching. The values of $\k2av_{pp} = $ 1.7 $GeV^2$, $\nu_m=4$,
and $C=0.4$~GeV$^2$ are fixed~\cite{zfpgl02} in the calculation.}
\end{figure}

The predicton on $K/\pi$ ratios at RHIC energy, $\sqrt{s}= 200$ GeV,
for both $d + Au$ and $Au + Au$ collisions, are shown in Fig.\ref{fig:fig3}.

In above calculations, we include parton multiscattering due to 
the effect of nuclear medium. The extra contribution to the width 
can be related to the number of nucleon-nucleon ($NN$) collisions  
in the medium as follows:
\begin{equation}
\label{ktbroadpA}
\k2av_{pA} = \k2av_{pp} + C \cdot h^{sat}_{pA}(b) \ .
\end{equation}
Here $\k2av_{pp}$ is the width of the transverse momentum distribution
of partons in $pp$ collisions and $\k2av_{pp} = $ 1.7 $GeV^2$ at 
$\sqrt{s} = 200$ GeV~\cite{zfpgl02}. The function $h_{pA}(b)$ describes the   
number of {\it effective} $NN$ collisions at impact parameter $b$ which  
impart an average transverse momentum squared $C$ and it can be written
in terms of the number of collisions suffered by the incoming proton in
the target nucleus, $\nu_A(b) = \sigma_{NN} t_A(b)$, where $\sigma_{NN}$ 
is the inelastic $NN$ cross section. In this work, we take the `saturated' 
prescription~\cite{plf00,zfpgl02}., i.e., we assume that only one 
associated $NN$ collision is responsible for the $\k2av$ enhancement, so 
$h_{pA}^{sat}(b)$ was equated to a maximum value $\nu_m$, whenever 
$\nu_A(b) \geq  \nu_m=2$. We fix the value of $\nu_m=4$ and an 
associated value of $C=0.4$~GeV$^2$\cite{zfpgl02}.
 
we also approximately include shadowing effect (the PDFs are modified
due to nuclear environment)~\cite{wang91,eskola99}
and the isospin asymmetry of heavy nuclei into the nuclear PDFs. 
We use the average nuclear dependence and the scale-independent
parameterization of the HIJING shadowing function~\cite{wang91}, 
\begin{eqnarray}
\label{shadowpdf}
 && f_{a/A}(x,Q^2) = R_{a/A}(x)\left[\frac{Z}{A}f_{a/p}(x,Q^2)\right]+R_{a/A}(x) \, \nonumber \\
 && \ \ \ \ \ \ \ \ \ \ \ \ \ \ \ \, \times \left[\left(1-\frac{Z}{A}\right)\, f_{a/n}(x,Q^2) \right],
\end{eqnarray}
where $f_{a/n}(x,Q^2)$ is the PDF for the neutron. 

In conclusion, an improved LO FFs set of kaon production is proposed 
in present letter, based on the experimental data. Relying on this 
improved FFs set, we calculate the 
high-$p_T$ $K/\pi$ ratios in $pp$ collisions in center-of-mass energy
range 20 GeV $\lesssim \sqrt{s} \lesssim$ 40~GeV, and we find our 
results in $pp$ collisions are in good agreement with experiment data. 
We also predict
$K/\pi$ ratios for $d + Au$ and $Au + Au$ collisions at $\sqrt{s} = 200$ 
GeV to provide background information for further comparison with the 
high-$p_T$ kaon production data of RHIC. The future systematic investigation 
of kaon production with this improved FFs set in $pp$, $pA$ and $AB$ collisions will be
very interesting and be reported in other place. For $AB$ collisions with the 
consideration of
jet quenching, the current FFs set of kaon will give very important insight  
about the existence and properties of the QGP. 

\vspace*{4 mm}
The author thank to Dr. Fai and Dr. Kretzer
for useful comments and stimulating discussions. 

\vspace*{4 mm}

\end{document}